\documentclass{PoS}

\usepackage[latin1]{inputenc}
\usepackage{graphicx}
\usepackage{epsfig}
\usepackage{subfigure}
\usepackage{latexsym}
\usepackage{amsmath}

\newcommand{\ev}[1]{\langle #1 \rangle}

\newcommand{\csw}{c_{\rm{sw}}}

\newcommand{\be}{\begin{equation}}
\newcommand{\ee}{\end{equation}}
\newcommand{\bea}{\begin{eqnarray}} 
\newcommand{\eea}{\end{eqnarray}}

\newcommand{\bmp}{\noindent\begin{minipage}{16cm}}
\newcommand{\emp}{\end{minipage}\vskip 7mm} 
\def\lsim{\mathrel{\raise.3ex\hbox{$<$\kern-.75em\lower1ex\hbox{$\sim$}}}}
\def\gsim{\mathrel{\raise.3ex\hbox{$>$\kern-.75em\lower1ex\hbox{$\sim$}}}}

\newcommand{\intron}[1]{}

\title{Exploring the conformal window: SU(2) gauge theory on the lattice}

\ShortTitle{SU(2) gauge theory on the lattice}

\author{\speaker{Tuomas Karavirta}\\
	Department of Physics, P.O.Box 35 (YFL), 
        \\ FI-40014 University of Jyv\"askyl\"a, Finland, 
        \\ and 
  	    \\ Helsinki Institute of Physics, P.O.~Box 64, 
  	    \\ FI-00014 University of Helsinki, Finland\\
	E-mail: \email{tuomas.karavirta@jyu.fi}}

\author{Jarno Rantaharju\\
	Department of Physics and Helsinki Institute of Physics,\\
	P.O.Box 64, FI-00014 University of Helsinki, Finland\\
	Email: \email{jarno.rantaharju@helsinki.fi}}

\author{Kari Rummukainen\\
 	Department of Physics and Helsinki Institute of Physics,\\
 	P.O.Box 64, FI-00014 University of Helsinki, Finland\\
	Email: \email{kari.rummukainen@helsinki.fi}}

\author{Kimmo Tuominen\\
	Department of Physics, P.O.Box 35 (YFL), 
        \\ FI-40014 University of Jyv\"askyl\"a, Finland, 
        \\ and 
  	    \\ Helsinki Institute of Physics, P.O.~Box 64, 
  	    \\ FI-00014 University of Helsinki, Finland\\
	Email: \email{kimmo.i.tuominen@jyu.fi}}

\abstract{We study the SU(2) gauge theory on the lattice with different numbers of fermions in the fundamental representation of the gauge group to explore the gauge theory phase diagram. We find evidence for an infrared fixed point for ten flavors. The theory with six flavors shows behaviors compatible with the existence of a (quasi) stable fixed point, but the large errors in the present data do not allow for decisive confirmation of this.} 

\FullConference{The XXIX International Symposium on Lattice Field Theory, Lattice2011\\
		July , 2011\\
		}

\begin{document}

\section{Introduction}
There exists a class of gauge theories where, under the renormalization group evolution, the coupling 
shows asymptotic freedom at small distances, analogously to QCD, but flows to a fixed
point at large distances where the theory hence looks conformal. Such theories have
applications in beyond Standard Model model building. These include
unparticles, i.e. an infrared conformal sector coupled weakly to the
Standard Model \cite{Georgi:2007ek}, and
(extended) technicolor scenarios, that explain the masses of the
Standard Model gauge bosons and fermions via strong coupling gauge
theory dynamics \cite{TC,Hill:2002ap,Sannino:2008ha}. 
In addition to direct applications to particle phenomenology, the phase diagrams of 
gauge theories, as a function of the number of colours, $N$, flavours $N_f$ and fermion 
representations, are interesting from the purely theoretical viewpoint of understanding the nonperturbative gauge theory dynamics from first principles. While several semianalyitc methods to estimate the vacuum phase diagram of a gauge theory exist, the only truly first principle method is constituted by lattice simulations. Several initial studies have appeared in literature: for example SU(2) with fundamental representation fermions \cite{Bursa:2010xr}, SU(2) with adjoint fermions \cite{Catterall:2007yx,Hietanen:2008mr,DelDebbio:2008zf,Catterall:2008qk,Hietanen:2009az,DeGrand:2011qd} and SU(3) with fermions in the fundamental  \cite{Appelquist:2007hu,Fodor:2009wk,Deuzeman:2008sc} or in the two-index symmetric \cite{Shamir:2008pb}, i.e. the sextet, representation.

The studies with Wilson fermions are subject to lattice artifacts proportional to the lattice spacing $a$, and a program to cancel these lattice artifacts has been devised \cite{Sheikholeslami:1985ij,Luscher:1992an}. As a motivation for this improvement, 
consider the measurement of the running coupling using the Schr\"odinger functional method:
The coupling is measured using a background field and the scale is set by the finite size of the 
lattice. We consider a lattice of volume $V=L^4=(Na)^4$. The spatial links at the 
$t=0$ and $t=L$ boundaries are fixed to constant values,
while the spatial boundary conditions are periodic.
The fermion fields are set to vanish at the $t=0$ and $t=L$ boundaries and have twisted 
periodic boundary conditions in spatial directions: 
$\psi(x+L\hat i) = \exp(i\pi/5)\psi(x)$.
At the classical level, the boundary conditions generate a constant chromoelectric 
field and the derivative of the action with respect to $\eta$ can be easily calculated:
\begin{align}
\frac{\partial S^{\textrm{cl.}}}{\partial \eta} = \frac{k}{g^2_0},
\end{align}
where $k$ is a function of $N=L/a$ and $\eta$.
At the full quantum
level the coupling is defined by
\begin{align}
\ev{ \frac{\partial S}{\partial \eta} } = \frac{k}{g^2}.
\end{align}
The perturbative step scaling function defined using the evolution of the 
renormalized coupling $g$ from
scale $L$ to scale $sL$, i.e.
\bea
\Sigma(u,s,L/a) &=& g^2(g_0,sL/a)\vert_{g^2(g_0,L/a)=u} \nonumber \\
&=& u+(\Sigma_{1,0}+\Sigma_{1,1} N_f)u^2.
\label{Step scaling formula}
\eea
The second line gives the formula in perturbation theory to one loop order, and 
the fermion contribution is denoted by $\Sigma_{1,1}$. To evaluate these 
perturbative contributions we use the methods in 
\cite{Sint:1995ch,Sommer:1997jg}, and choose $s=2$. The continuum limit of $\Sigma_{1,1}$ is given by the fermionic contribution to the one loop coefficient $b_{0}=\beta_0/(16\pi^2)$ of the beta function, i.e. 
\be
\delta=\lim_{L/a\rightarrow 0}\Sigma_{1,1}/(2N_f b_{0,1}\ln{2}),
\ee
where $b_{0,1}=1/(24\pi^2)$.

\begin{figure}
\centering
\includegraphics[scale=0.35]{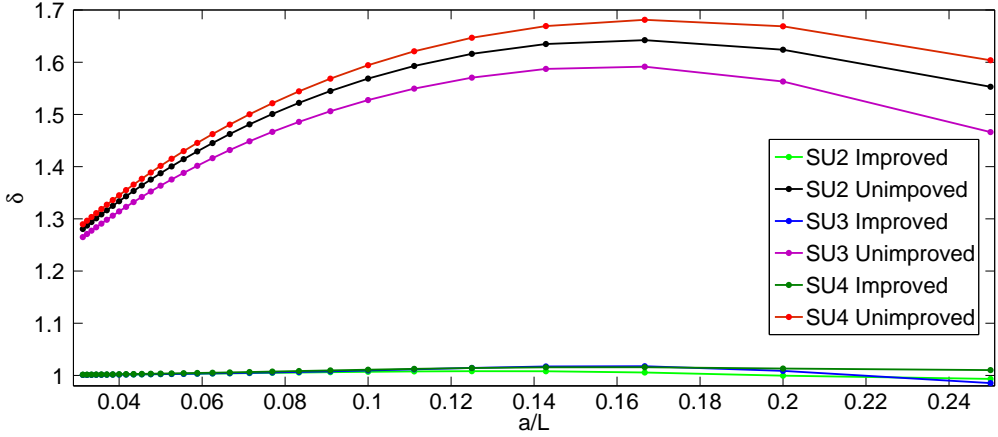}
\caption{Contribution of a massless Wilson quark to the step scaling function normalized to its continuum value at one loop order in perturbation theory. The top three curves show the result for gauge groups SU(4), SU(3) and SU(2) (from top to bottom) for unimproved Wilson fermions, while the lower three curves show the result after ${\cal{O}}(a)$ improvement has been taken into account.} 
\label{pertstep}
\end{figure}

The
results for the one loop fermion contribution is shown in figure \ref{pertstep} both 
for unimproved Wilson fermions and with
${\cal{O}}(a)$ improvement. One immediately observes that without improvement,
$\Sigma_{1,1}$ depends strongly on $L/a$ and approaches the continuum limit 
only for large lattices, while with improvement the large lattice artefacts are 
absent. Clearly this motivates the need to use improved 
actions in the lattice studies of these theories with Wilson fermions.

In our nonperturbative study we use Symanzik improved Wilson fermions to remedy the discretization errors. We measure
the running coupling in both models and the mass anomalous dimension in the model
with six fundamental fermions. See also \cite{Karavirta:2011zg}.

\section{The model and theoretical tools}
In this section we will introduce the model and some theoretical tools regarding the step scaling function and anomalous dimension of the mass operator. 
We use the basic Wilson Lattice action
\begin{eqnarray}
	S_{0}=S_G+S_F,
	\label{eq:aktion}
\end{eqnarray}
where $S_G$ is the standard Wilson plaquette action and
and $S_F$ is the clover improved Wilson action
\begin{equation}
  S_F
  = a^4\sum_{\alpha=1}^{N_f} \sum_x \left [
  \bar{\psi}_\alpha(x) ( i D + m_0 )
  \psi_\alpha(x)
   + a \csw \bar\psi_\alpha(x)\frac{i}{4}\sigma_{\mu\nu}
  F_{\mu\nu}(x)\psi_\alpha(x) \right ],
\end{equation}
where $D$ is the standard Wilson-Dirac derivative operator including
the doubler term.
We set the improvement coefficient $\csw$ to the perturbative value \cite{Luscher:1996vw}
$ \csw = 1 + 0.1551(1) g_0^2 + O(g_0^4)$.
We have performed a few short measurements that imply that at strong coupling this is close
to the correct nonperturbative value for $\csw$ with 6 and 10 fermion flavours.
We also include the perturbative improvement at the Schr\"odinger functional 
boundaries as described in \cite{Karavirta:2011mv}.

There is a relation between step scaling funtion \eqref{Step scaling formula} and the $\beta$ function:
\begin{eqnarray}
-2\ln(s)&=&\int_u ^{\sigma(u,s)}\frac{dx}{\sqrt{x}\beta({\sqrt{x}})}.
\label{Step & beta}
\end{eqnarray}
Near the fixed point $\beta$-function is small and 
\eqref{Step & beta} can be approximated with
\begin{eqnarray}
\beta(g) &\approx&\frac{g}{2\ln(2)} 
\left ( 1 - \frac{\sigma(g^2,s)}{g^2} \right ).
\label{beta approx}
\end{eqnarray}
We also measure the mass anomalous dimension $\gamma=d\ln m_q/d\ln\mu$ of the 
theory with 6 fermion flavours 
using the pseudoscalar 
density renormalization constant which is defined as
\begin{equation}
Z_P(L)=\frac{\sqrt{3} f_1}{f_P(L/2)},
\end{equation}
where 
\begin{equation}
f_1=\frac{-1}{12 L^6} \int d^3u d^3v d^3y d^3z 
  \langle\bar \zeta'(u)\gamma_5\lambda^a\zeta'(v)
  \bar\zeta(y)\gamma_5\lambda^a\zeta(z)\rangle,
\end{equation}
\begin{equation}
f_P(x_0)=\frac{-1}{12 L^6} \int d^3y d^3z 
  \langle\bar \psi(x_0)\gamma_5\lambda^a\psi(x_0)
  \bar\zeta(y)\gamma_5\lambda^a\zeta(z)\rangle,
\end{equation}
are correlation functions of the pseudoscalar density. 
Here sources $\zeta$ and $\zeta'$ are located at the $t=0$ and $t=L$ boundaries, respectively.
For these measurements the boundary matrices at $t=0$ and $t=L$ are set to unity.
The mass step scaling function is then defined as:
\begin{align}
 &\Sigma_P(u,s,L/a) = 
    \left. \frac {Z_P(g_0,sL/a)}{Z_P(g_0,L/a)} \right |_{g^2(g_0,L/a)=u} \quad,\\
 &\sigma_P(u,s) = \lim_{a/L\rightarrow 0} \Sigma_P(u,s,L/a),
\end{align}
and we choose again $s=2$. We find the continuum step scaling function $\sigma_P$ by
measuring $\Sigma_P$ at $L/a=6$ and $10$, and doing a quadratic extrapolation.  It can be related to the anomalous dimension of the mass operator by
\begin{equation}
 \sigma_P(u,s) = \left ( \frac{u}{\sigma(u,s)} \right ) ^{d_0/(2b_0)}
   \exp \left [\int_{\sqrt u}^{\sqrt{\sigma(u,s)}} dx
   \left ( \frac{\gamma(x)}{\beta(x)} - \frac{d_0}{b_0 x} \right )   \right ],
   \label{massstep}
\end{equation}
where $b_0=\beta_0/(16\pi^2)$ in terms of the one-loop coefficient $\beta_0$ of the beta function and 
$d_0=8/(16\pi^2)$ is the corresponding one-loop coefficient for the anomalous dimension, 
$\gamma=-d_0 g^2$.
This can be approximated at the fixed point with
$
 \gamma^\ast(g^2) = -\frac{\sigma_P(g^2,s)}{\log(s)}.
$

\section{Measurements and results}
The zero mass limit is determined by measuring the $\kappa_c =1/(8+2m_{0,c})$ for all the used values of $\beta$ via the PCAC relation using lattice size $16^4$. The measured values of $\kappa_c$ are then used for all lattice sizes.
The running coupling $g$ is measured for $\beta=\{8,6,4,3,2,1.7,1.5,1.3,1\}$ for $N_f=10$ and $\beta=\{8,5,4,3,2.4,2,1.8,1.5,1.44,1.4,1.39\}$ for $N_f=6$, using lattice sizes $6^4$, $8^4$, $12^4$ and $16^4$. The measured values of $g$ are then used to find an interpolating function in $\beta$ of the form
\begin{equation}
  \frac{1}{g^2(\beta,L/a)}=\frac{\beta}{2N} \sum_{i=0}^n c_i \left(\frac{2N}{\beta}\right)^i
\end{equation}
with $c_0=1$.
The interpolating function is used to find the step scaling function for $L/a=6,8$, and the continuum limit is extracted using 
\begin{equation}
\Sigma(u,2,L/a)=\sigma(u,2)+c\left(L/a\right)^2.
\end{equation}
Because of the improved action we expect the $\mathcal{O}(a)$ terms to be subleading.  Unfortunately, with only two points in the extrapolation, it is not possible to verify the accuracy of the extrapolation quantitatively.

The anomalous dimension of the mass operator is determined similarly: The pseudoscalar density renormalization constant $Z_P$ is measured for $\beta=\{2.4,2,1.5,1.44,1.4,1.39\}$ and for $L/a=6,8,10,12,16,20$, and these values are used to find an interpolating function of the form 
\begin{equation}
Z_P(\beta,L/a)=\sum_{i=0}^n c_i \left(\frac{1}{\beta}\right)^i
\end{equation}
with $c_0=1$.
This is then used to calculate $\Sigma_P(u,2,L/a)$, which is in turn 
extrapolated to the continuum limit with an order $(a/L)^2$ ansatz.

\begin{figure}
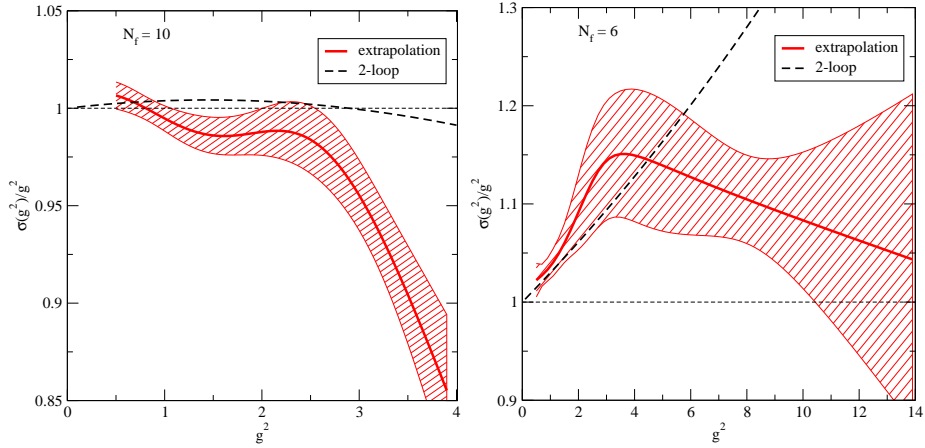

\centering
\includegraphics[width=0.4\textwidth]{nf10.eps}
\includegraphics[width=0.4\textwidth]{nf6.eps}
\caption{Step scaling function for $N_f=10$ (left) and $N_f=6$ (right).
Continuous lines show 2-loop perturbative step scaling result.} 
\label{Step scaling figure}
\end{figure}

In figure \ref{Step scaling figure} we show the step scaling functions. 
In the $N_f=10$ theory the evolution of the coupling is extremely slow, and our results basically agree with this at $g^2\lsim 2.5$: the step scaling practically vanishes in this range.  In this case we expect the two-loop perturbative step scaling function to be fairly accurate, and from figure \ref{Step scaling figure} we see that the errors should be an order of magnitude smaller in order to resolve it.  At stronger coupling the measured step scaling deviates significantly from zero to negative values.  Combined with the analytically known weak coupling behaviour, this indicates that the $\beta$-function must have a fixed point somewhere in this range.   
However, we believe that a large fraction of the observed deviation from the perturbative step scaling at strong coupling arises from the results at our strongest lattice coupling $\beta_L = 4/g_0^2 = 1$.  This point deviates clearly from the rest of the simulation points, possibly indicating stronger cutoff effects.


In the $N_f=6$ theory the evolution of the coupling remains slow, which leads to large errors in the step scaling function.  In this case we were able to reach mesured couplings up to $g^2 = 14$.
However, the results indicate that the possible infrared fixed point is at $g^2 \gsim 13$, and our statistical resolution is not sufficient to confirm or exclude the existance of an IRFP.



\begin{figure}
\centering
\includegraphics[scale=0.4]{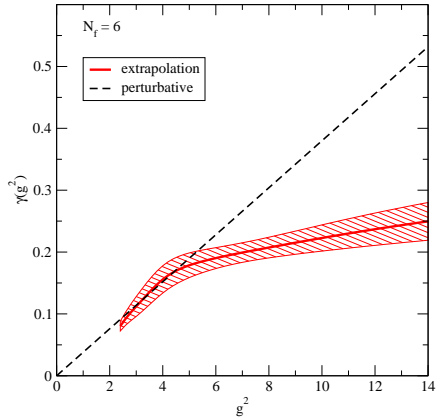}
\caption{Anomalous dimension of the mass operator for $N_f=6$} 
\label{gamma}
\end{figure}

The measured values of mass anomalous exponent $\gamma$ are shown in figure \ref{gamma}. It shows that $0.1<\gamma<0.3$. Because the value of the anomalous dimension of the mass operator is only scheme independent at the fixed point, these results are not that interesting, since we were unable to find the IRFP. 

\section{Conclusions}

Our simulations verify that the SU(2) gauge theory with 6 flavours of fundamental representation fermions is indeed close to the lower edge of the conformal window.  Unfortunately, the 
possible fixed point in this theory is at such a strong coupling that we were not able to fully resolve the behaviour: the results are compatible either with a fixed point at $g^2 \gsim 12$ or with a ``walking'' behaviour where the $\beta$-function almost vanishes.  The value of the fixed point coupling is naturally scheme dependent; this value is for Schr\"odinger functional scheme.  To resolve this question requires simulations with an action which can be used at stronger lattice couplings than used in this work.

\acknowledgments
T.K. is supported by the Magnus Ehrnrooth foundation and by University of Jyv\"askyl\"a Faculty of Mathematics and Science.  J.R. is supported by the Finnish Academy of Science and Letters V\"ais\"al\"a fund.  We acknowledge the support from the Academy of Finland grant
number 1134018.  The computations have been performed at the Finnish IT Center for Science.

\end{document}